\documentstyle[prb,psfig,aps,floats]{revtex}

\newcommand{\be}{\begin{equation}} 
\newcommand{\ee}{\end{equation}}
\newcommand{\bt}{\begin{tabular}} 
\newcommand{\et}{\end{tabular}}

\begin{document}

\twocolumn[\hsize\textwidth\columnwidth\hsize\csname@twocolumnfalse%
\endcsname

\draft

\title{Dynamical Critical Properties of the Random Transverse-Field Ising 
Spin Chain}

\author{J. Kisker} \address{Institut f\"ur Theoretische Physik,
  Universit\"at zu K\"oln, D-50937 K\"oln, Germany}

\author{A. P. Young} \address{Department of Physics, University of California,
  Santa Cruz, CA 95064}

\date{\today}

\maketitle

\begin{abstract}
We study the dynamical properties of the random transverse-field 
Ising chain at criticality
using a mapping to free fermions, with which we can obtain
numerically exact results for system sizes, $L$, as large as 256.
The probability distribution of the local imaginary time correlation 
function $S(\tau)$ is investigated and found to be simply a function of
$\alpha  \equiv -\log S(\tau) / \log \tau$.
This scaling behavior implies that the {\em typical}
correlation function decays
algebraically, $S_{{\rm typ}}(\tau) \sim \tau^{-\alpha_{\rm typ}}$,
where the exponent $\alpha_{\rm typ}$
is determined from 
$P(\alpha)$, the distribution of $\alpha$.
The precise value for $\alpha_{\rm typ}$ depends on exactly how the
``typical'' correlation function is defined. The form of $P(\alpha)$ for 
small $\alpha$ gives a contribution to the {\em average} correlation function,
$S_{{\rm av}}(\tau)$, namely
$S_{{\rm av}}(\tau) \sim(\log \tau)^{-2x_m}$, where
$x_m$ is the bulk magnetization exponent, which was obtained recently in
Europhys. Lett. {\bf 39}, 135 (1997). These results represent a type of
``multiscaling'' different from the well-known
``multifractal'' behavior.

\end{abstract}

\pacs{}

\vskip 0.1 truein
]

\section{Introduction}
Quantum phase transitions show a number of remarkable features and have 
attracted considerable interest in recent years. They occur at zero
temperature and are driven by quantum, rather than thermal, fluctuations.
Hence, they are induced by varying a parameter other than the
temperature, such as an applied transverse magnetic field. In particular,
systems with quenched disorder show surprising properties
near a quantum critical point. 
For instance, it was found that in quantum Ising spin glasses 
\cite{RY_qsg,GBH} and random transverse-field Ising ferromagnets,
 \cite{Fisher,SS} as well as in ``Bose glass'' systems
(which have a continuous symmetry of the order parameter but lack
``particle-hole'' symmetry \cite{KR}), 
all or part of the disordered phase shows features which are usually 
characteristic of a critical point. More precisely, correlations in time decay 
algebraically and thus various susceptibilities may actually diverge.
This behavior is due to Griffiths-McCoy singularities 
\cite{Griffiths,McCoy}, which arise from rare clusters which 
are more strongly coupled than the average, and it has become clear
that their effect is much more pronounced near a quantum transition than 
near a classical critical point.

One of the simplest models exhibiting the characteristic features of a 
quantum phase transition is the random 
transverse-field Ising chain, defined by the Hamiltonian
\be
H=-\sum_{i=1}^{L-1}J_i\sigma_i^z\sigma_{i+1}^z-\sum_{i=1}^L h_i\sigma_i^x,
\label{hamil}
\ee
where the $\{\sigma_i^\alpha\}$ are Pauli spin matrices at site $i$ and
the interactions $J_i$ and the transverse fields $h_i$ are random
variables with distributions $\pi(J)$ and $\rho(h)$, respectively.

A lot of results on the critical and off-critical properties of this model 
have been obtained, both analytically and numerically. The ground state 
properties of the Hamiltonian in Eq. (\ref{hamil}) are closely related
to a two-dimensional classical Ising model where the disorder is perfectly
correlated along one direction, the latter model first being studied 
by McCoy and Wu \cite{MW}. Subsequently, the quantum model was studied
by Shankar and Murthy \cite{SM}, and recently the critical properties
have been worked out in great detail by D.~S. Fisher, using a real space
renormalization group approach \cite{Fisher,FY}. The quantum model, 
Eq. (\ref{hamil}), has also been 
investigated numerically \cite{FY,IR,APY,RI_epl,YR}
using a mapping to free fermions by means of a 
Jordan-Wigner transformation.

We now briefly summarize some of the surprising features of this model.
Distributions of {\em equal time}
correlation functions are found to be very broad
which leads to \cite{Fisher}:
(i) different critical
exponents for the divergence of the ``typical'' \cite{typical} and average
correlation lengths, and (ii) the typical equal time correlation function at
criticality falls off as a stretched exponential function of distance, quite
different from the power law variation of the average.
Recently, it has been shown that these main results are not restricted to
one dimension, but also seem to
hold in the two-dimensional random Ising ferromagnet
\cite{PY,RK}. 

A number of results for
dynamics have also been found. For example, at the critical point the
dynamical exponent $z$ is infinite \cite{Fisher}.
{\em Away from the critical point} the
distribution of local (imaginary) time dependent correlation functions is very
broad \cite{APY}.
The average varies as a (continuously varying) power of imaginary
time $\tau$
involving an exponent \cite{zprime} $z^\prime(\delta)$, where $\delta$ is the
deviation from criticality. By contrast, the typical correlation function
varies as a stretched exponential function of time. Quite detailed information
on the whole distribution of time dependent correlation functions in the
paramagnetic phase has also been
found \cite{APY}.
{\em At the critical point} the average correlation function is found
to decay with an inverse power of the log of the time \cite{RI_epl},
corresponding to the
result $z=\infty$ mentioned above.

However, the {\em distribution} of
time dependent correlation functions at the critical point
has not yet been determined and is the focus of this
study. We find that the distribution is very broad and
can be expressed in terms of a single (logarithmic)
scaling variable, defined in Eq.~(\ref{alpha}) below. This result implies
that there is a continuous range of exponents $\alpha$, somewhat analogous to
(but also with important differences from) 
``multifractal'' behavior \cite{halsey,Ludwig}.

We present our results in the following section, and we conclude with a 
discussion in Sec. \ref{sec_discussion}.

\section{Results}
\label{sec_results}
Throughout the paper we assume the following rectangular distributions
for the couplings $J_i$ and the transverse fields $h_i$
\begin{eqnarray}
\pi (J) & = & \left\{ \begin{array}{ll} 1 & \mbox{for} \quad 0 < J < 1\\ 
                                   0 & \mbox{otherwise}
                  \end{array} \right.\\
\rho (h) & = & \left\{ \begin{array}{ll} h_0^{-1} & \mbox{for} \quad 
    0 < h <h_0\\
    0 &\mbox{otherwise,} \end{array} \right.
\end{eqnarray}
which are characterized by a single control parameter $h_0$.
The system possesses a critical point at 
$\delta=[\ln J]_{{\rm av}} - [\ln h]_{{\rm av}}=0$, 
i.e. $h_0=1$,
at which the distributions of the bonds and fields are equal.
The lattice size is $L$ and we impose {\em free} rather than the more
conventional periodic boundary conditions \cite{IR,APY}.

For the numerical work we make use of the mapping of Hamiltonian  
(\ref{hamil}) onto a model of free fermions \cite{LSM,Katsura,Pfeuty}. 
Since the transformation has been used in previous work, we only give a brief 
summary here and refer to Refs. \cite{IR,APY,YR} for further details.
For free boundary conditions, which we shall assume here, the most convenient 
representation is given in Refs. \cite{IR,IT}, necessitating
only the diagonalization of a $2L\times2L$ real, {\em tridiagonal} matrix. 
The spin operators occurring in the expectation value of the correlation 
functions can then be expressed as a product of fermion operators, which is 
evaluated using Wick's theorem. The resulting Pfaffian is then given by the 
square root of the determinant of a matrix, where the matrix elements can be 
calculated from the eigenvectors and eigenvalues of the Hamiltonian in the 
free fermion representation. The imaginary time correlation functions are
always positive, so there is no ambiguity in sign when taking the square root.
For $L \le 128$ we average over 30000 realizations of the disorder, while for
the largest size, $L=256$, we average over 10000 realizations.

We calculate the probability distribution $P(\ln S(\tau))$ of 
the single site imaginary time correlation function 
\be
S_{ii}(\tau)=\langle \sigma_i^z(\tau)\sigma_i^z(0) \rangle
\ee
at the critical point, i.e. $\delta=0$. For convenience, we will denote
$S_{ii}(\tau)$ by $S(\tau)$ from now on. To obtain better statistics, 
we determine the correlation function for
every second site from $L/4$ to $L/2$, making a total of $L/8$ sites.
All sites are far from the boundary so we do not expect the results to be
affected by boundary effects.
The average correlation function is then given by
\be
S_{\rm av}(\tau)=\frac{8}{L} \sum_i [S_{ii}(\tau)]_{\rm av} ,
\ee
where $[\cdots]_{\rm av}$ denotes an average over samples.

\begin{figure}
\begin{center}
\leavevmode
\psfig{file=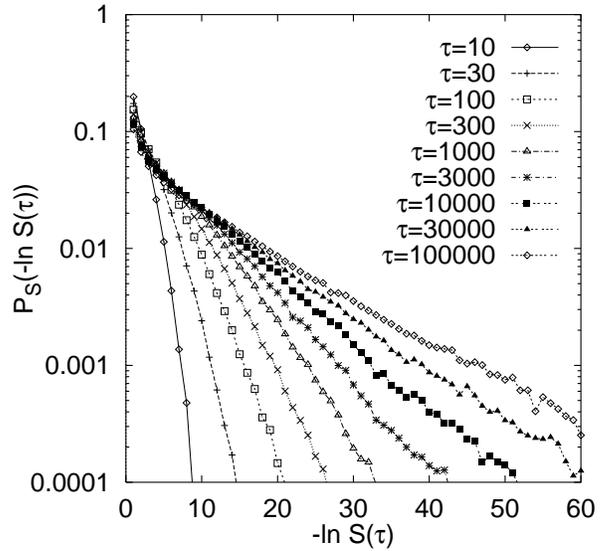,width=\columnwidth}
\end{center}
\caption{The probability distribution, $P_S$, of $-\ln S(\tau)$
for $L=128$ 
and different values of the imaginary time $\tau$ at the critical point 
$\delta=0$. The data is averaged over $30000$ samples.}
\label{fig1}
\end{figure} 

\begin{figure}
\begin{center}
\leavevmode
\psfig{file=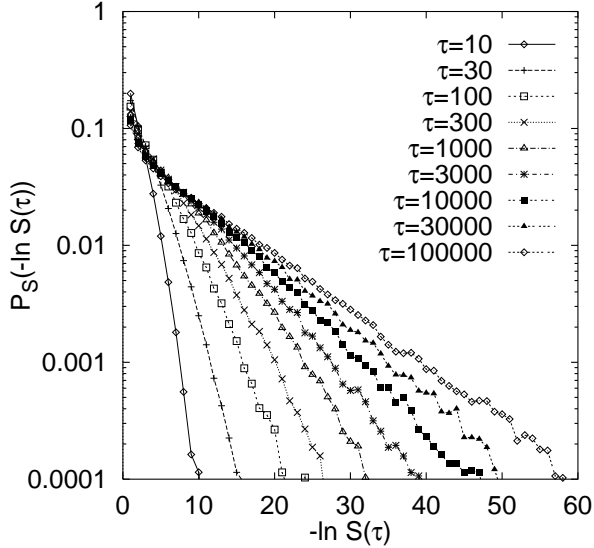,width=\columnwidth}
\end{center}
\caption{The probability distribution of $-\ln S(\tau)$
for $L=256$ 
at the critical point $\delta=0$. The disorder average is over 
10000 samples.}
\label{fig2}
\end{figure} 

Since we expect strong finite size effects at the critical point, we
calculated data for different system sizes. Data for
the distribution of $-\ln S(\tau)$ for
$L=128$ and $L=256$ is
shown in Figs. \ref{fig1} and \ref{fig2}.
One observes that the distributions are broad
and that for larger times the probability distribution gains more 
weight in the tail, indicating that correlations decrease for larger times, as
expected.

Since the curves for fixed $L$ and different times $\tau$ appear to be
shifted by a roughly constant amount on the (logarithmic) x-axis,
we attempt a scaling plot of the data with the parameter free
scaling variable 
\be
\alpha=-\frac{\ln S(\tau)}{\ln \tau} \,.
\label{alpha}
\ee
Note that in Ref. \onlinecite{APY}, which investigated the dynamics in the
paramagnetic phase, 
the scaling variable was found
to be $-\ln S(\tau) / \tau^{1/\mu}$, where $\mu$ is expected to diverge at the
critical point.
Hence $\alpha$ in Eq. (\ref{alpha}) is a natural scaling 
variable at the critical point.

\begin{figure}
\begin{center}
\leavevmode
\psfig{file=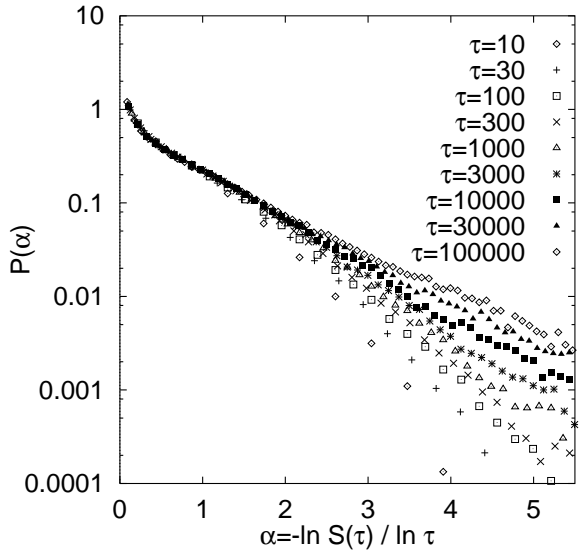,width=\columnwidth}
\end{center}
\caption{Scaling plot of the probability distribution in
  Fig. \ref{fig1} ($L=128$). The scaling variable $\alpha$ is that given in 
  Eq. (\ref{alpha}). For larger values of $\alpha$ systematic deviations from 
scaling occur. This comes from data for small times $\tau$ which is
presumably not in the scaling region.}

\label{fig3}
\end{figure} 

\begin{figure}
\begin{center}
\leavevmode
\psfig{file=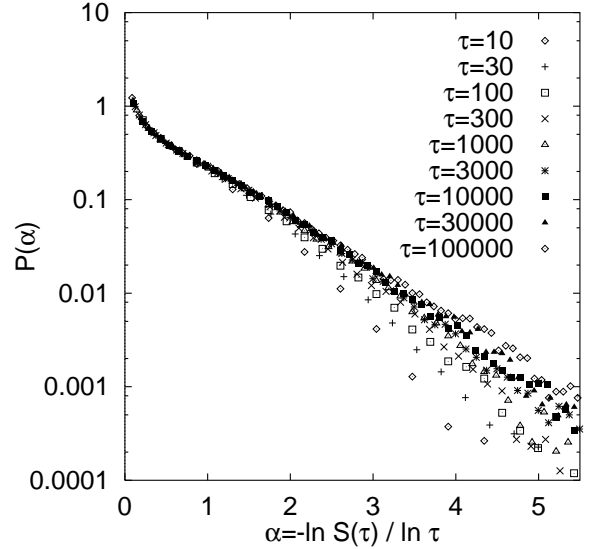,width=\columnwidth}
\end{center}
\caption{Scaling plot of the probability distribution in
  Fig. \ref{fig2} ($L=256$). The scaling variable $\alpha$ is that given in 
  Eq. (\ref{alpha}). Note, by comparison with Fig.~\ref{fig3}
  that the range where the data scale well increases
  with increasing system size.
}
\label{fig4}
\end{figure} 

The corresponding scaling plots for $P(\alpha)$ against
$\alpha$ are shown in Figs. \ref{fig3} and \ref{fig4} 
for $L=128$ and $L=256$.
One sees that the data collapse is good for $\alpha$ not too large 
and that the range of $\alpha$ where scaling works {\em increases} with
increasing $L$. The data for large $\alpha$ where scaling breaks down
corresponds to {\em small} $\tau$, and it is reasonable to expect deviations
from scaling in this region.

Since the data is only a function of the scaling variable $\alpha$, it
follows
that typically the correlation function falls off with a power of $\tau$. We
shall now see that the precise value of the power depends in detail on how the
typical correlation function is defined. For example,
if we define ``typical" to be the exponential of the average of the $\log$, 
i.e.
\be
S_{\rm avlog}(\tau) =\exp([\ln S(\tau)]_{\rm av}) \,,
\label{s_typ}
\ee
we obtain
\be
[\ln S(\tau)]_{{\rm av}} = -\int_0^\infty P(\alpha)\alpha \ln \tau \,  d\alpha
= -\langle \alpha \rangle \ln \tau \, ,
\ee
which yields,
the algebraic decay
\be
S_{\rm avlog}(\tau) = \tau^{-\langle \alpha \rangle}
\,,
\label{s_typ_alg}
\ee
where $\langle \ldots \rangle$ denotes an average with respect to the
distribution $P(\alpha)$. From our data we get
$\langle \alpha \rangle \simeq 0.7$. 

On the other hand if we define ``typical" to be the median of the 
distribution, then one easily sees that
\begin{equation}
S_{\rm median}(\tau) = \tau^{-\alpha_{\rm med} } ,
\end{equation}
where $\alpha_{\rm med}$ is the median of $P(\alpha)$, i.e. it is defined
implicitly by
\begin{equation}
{1\over 2} = \int_0^{\alpha_{\rm med}} P(\alpha) \, d\alpha .
\end{equation}

Any reasonable
definition of ``typical" will give a power law, in contrast to the
average which has a much slower logarithmic variation, which we discuss next.

Contributions to the {\em average} correlation function
can come both from the scaling
function and from non-scaling contributions \cite{YR,FY}.
The scaling part comes from
the small $\alpha$ part of the scaling function
where there is an upturn in the data, similar to that found for
the scaling function of the distribution of the static spin-spin
correlations $C(r) =\langle \sigma_i^z(0)\sigma_{i+r}^z(0) \rangle$ at the
critical point \cite{YR}. If we assume an 
algebraic relation $P(\alpha) \sim \alpha^{-\lambda}$ for 
small $\alpha$, we can calculate the scaling contribution to the
average correlation
$S_{\rm av}(\tau)$ from
\be
S_{\rm av}(\tau) = \int_0^\infty P(\alpha) S(\tau) \, d\alpha.
\ee
Noting that $S(\tau) = \exp( -\alpha \ln \tau)$ one obtains
\begin{eqnarray}
  S_{\rm av}(\tau) & \sim & \int_0^\infty \alpha^{-\lambda}
                            \exp(-\alpha \ln \tau)   \, d\alpha \nonumber\\
                   & \sim & (\ln \tau)^{-(1-\lambda)}.
\label{corrt_lambda}
\end{eqnarray}
This agrees with the results of Rieger and Igl\'oi \cite{RI_epl} who found
\be
S_{\rm av}(\tau) \sim (\ln \tau)^{-2x_m}\, ,
\label{corrt_av}
\ee
(where $x_m=(1-\phi/2) \simeq 0.191$ is the bulk 
magnetization exponent with $\phi=(1+\sqrt{5})/2$), provided
$\lambda = 1 - 2 x_m \simeq 0.618$.

\begin{figure}
\begin{center}
\leavevmode
\psfig{file=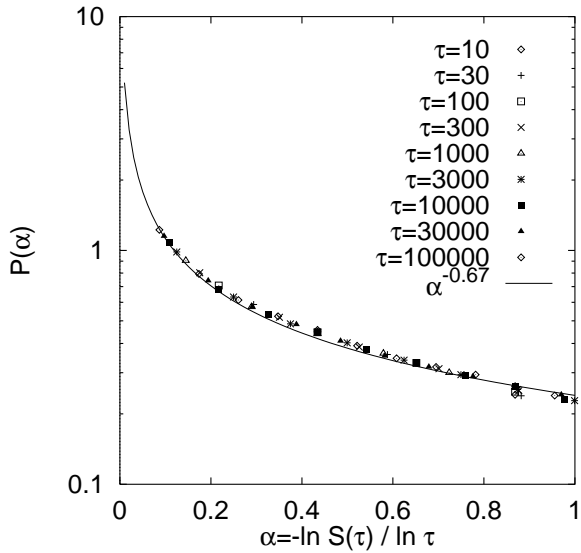,width=\columnwidth}
\end{center}
\caption{An enlarged plot of the data in Fig. \ref{fig4} (L=256).
  The full line is $\sim \alpha^{-0.67}$.
}
\label{fig5}
\end{figure}

To check this 
we show in Fig.~\ref{fig5} an enlarged plot of the data
in Fig. \ref{fig4} ($L=256$) for
small $\alpha$, together with the function 
$\sim\alpha^{-0.67}$, which is the best power law fit to
the data.
The curve fits
the data fairly well and the exponent of $0.67$
is reasonably close to the value of $\lambda\simeq0.618$ calculated above.
Note again that there may be {\em additional} non-scaling contributions to the
average correlation function as in Ref.~ \onlinecite{FY}.

\section{Conclusions}
\label{sec_discussion}
We have studied numerically the distribution of local (on-site)
correlations in imaginary time
for the random transverse-field Ising chain at the critical point. 
The distribution was found to be logarithmically broad and the scaling 
variable $\alpha = \log S(\tau) / \log \tau$ was established.
This means that while the correlations typically decay with a power
of $\tau$, there is a {\em range} of exponents $\alpha$ with a distribution 
$P(\alpha)$.
The small 
$\alpha$ part of the scaling function, which dominates the average
correlations, can be fitted by $P(\alpha) \sim \alpha^{-0.67}$, which
gives (close to)
the correct exponent in Eq.~(\ref{corrt_lambda}) for the logarithmic
decay of the average
correlation function.
Note that {\em all} positive moments are determined by the small $\alpha$ region
and so fall off with the {\em same}\/ \cite{not-simple}
decay given in Eq.~(\ref{corrt_lambda}).

The behavior of the distribution of $S(\tau)$ found here is somewhat analogous
to ``multifractal'' behavior \cite{halsey} predicted for the decay of 
{\em spatial} correlations in the {\em classical} two-dimensional Potts model
at the critical point \cite{Ludwig}, since both have a distribution of scaling
exponents. Beyond that, however, there are significant differences. Whereas in
our case the probability of having a scaling exponent $\alpha$ is $P(\alpha)$,
which does not explicitly depend on $\tau$, for the corresponding multifractal
behavior, the probability would be $\tau^{-f(\alpha)}$.  As a result, for
multifractal
behavior, the typical correlation function has an exponent $\alpha_{\rm min}$,
the value of $\alpha$ at the minimum of
$f(\alpha)$, whereas here we find that the exponent for the typical
correlation function depends on exactly how ``typical'' is defined. Averages
of the $n$-th moment of the correlation function for positive $n$ are also
quite different. In our case, for all $n>0$, the behavior is dominated by the
small $\alpha$ region of the distribution and the exponent is independent of
$n$ \cite{not-simple}, whereas for multifractal behavior, the moments depend
on $n$ in a non-trivial way and are given by \cite{halsey,Ludwig}
the Legendre transform of
$f(\alpha)$. It would be interesting to see if there are other systems which
have a ``multiscaling'' behavior of the type found here. 

\section{Acknowledgments}
J.K. is indebted to H. Rieger for valuable hints on the numerics and useful
discussions. He thanks the Department of Physics of UCSC for its kind
hospitality and the Deutsche Forschungsgemeinschaft (DFG) for financial
support. This work is supported by NSF grant DMR 9713977.


\begin{references}
\bibitem{not-simple}
  i.e. the $n$-th moment, $X_n$ say, does not depend on
  $n$; this is different from the simple behavior of non-random or
  weakly-random systems where $X_n \propto n$.
\end{references}
\end{document}